\documentstyle[amsfonts,aps,prl,preprint]{revtex}
\begin{document}
\draft
\title{Critical State Theory for Non-Parallel Flux Line Lattices in Type-II Superconductors}

\author{A. Bad\'{\i}a$^{1}$ and C. L\'opez$^{2}$}
\address{$^{1}$Depto. de F\'{\i}sica de la Materia Condensada-I.C.M.A., $^{2}$Depto. de Matem\'atica Aplicada,
\\C.P.S.U.Z., Mar\'{\i}a de Luna 3, E-50.015
Zaragoza (Spain)}
\date{\today}
\maketitle
\vspace{5mm}
\begin{abstract} 
Coarse-grained flux density profiles in type-II superconductors with non-parallel vortex configurations are obtained by a proposed phenomenological least action principle. We introduce a functional ${\cal C}[\vec{H}]$ which is minimized under a constraint of the kind $\vec{J}\in\Delta (\vec{H},\vec{x})$, where $\Delta$ is a bounded set. In particular, we choose the isotropic case $|\vec{J}|\leq J_c(H)$, for which the field penetration profiles $\vec{H}(\vec{x},t)$ are derived when a changing external excitation is applied. Faraday's law, and the principle of minimum entropy production rate for stationary thermodynamic processes dictate the evolution of the system. Calculations based on the model can reproduce the physical phenomena of flux transport and consumption, and the striking effect of magnetization collapse in crossed field measurements.
\end{abstract}
\vspace{10mm}
\narrowtext
%
%%%%%%%%%%%%%%%%%%%%%%%%%%%%%%%%%%%%%%%%%%%%%%%%%%%%%%%%%%%%%%%%%%%%%%%%%%%%%%%
%%
%%
%%%%%%%%%%%%%%%%%%%%%%%%%%%%%%%%%%%%%%%%%%%%%%%%%%%%%%%%%%%%%%%%%%%%%%%%%%%%%%%
%
The magnetization curve of type-II superconductors may display physical properties against the expectations of equilibrium thermodynamics. In particular, the existence of hysteresis and non-negative magnetic moment has been routinely observed. The Critical State Model (CSM), which dates back to the work by C. P. Bean\cite{bean}, has been an essential phenomenological framework for the interpretation of the aforementioned experimental facts. The following prescription was given: {\em External field variations are opposed by the maximum current density $J_c$ within the material. After the changes occur $J_c$ persists in those regions which have been affected by an electric field}.

Currently, the irreversible properties of superconductors are well understood in terms of the vortex flux line lattice (FLL) dynamics in the presence of pinning centers. Within the framework of self-organized extended dynamical systems one can conceive the CSM as the competition between a repulsive vortex-vortex interaction and attractive forces towards the pinning centers\cite{richardson}. This results in metastable equilibrium states for which the gradient in the density of vortices is maximum, corresponding to the critical value for the macroscopic current density $J_c$. At the macroscopic level, one usually makes the assumption that the rearrangement to new equilibrium states is instantaneous whenever the system is perturbed. As a matter of fact, when the FLL is unpinned by an external drive, a diffusion process is initiated, which is characterized by a time constant 
$\tau_{\rm f} \sim \mu_{0}L^{2}/\rho_{\rm f}$ ($\rho_{\rm f}$ stands for the flux flow resistivity and $L$ is some typical length of the sample). Thus, the previous hypothesis corresponds to neglecting $\tau_{\rm f}$ as compared to the excitation typical period.

A serious limitation of Bean's model is that one can just apply it to lattices of parallel flux tubes. However, a wealth of experimental phenomena is related to interactions between twisted flux lines. Macroscopically, if vortex crossing is present, $\vec{J}$ develops full vectorial character and the condition $| \vec{J}| =J_c$ (or 0) does not suffice. To the moment, several phenomenological theories are at hand, which allow to deal with such cases. Among them we want to detail the work by Clem and P\'erez-Gonz\'alez\cite{clemgonzalez}. These authors have developed a model (double critical state model or DCSM in what follows) which includes current density components perpendicular and parallel to the local magnetic induction $\vec{B}$. Their corresponding critical values $J_{c\perp}$ and $J_{c\|}$ are respectively associated to depinning and flux-cutting phenomena. Metastable distributions of $\vec{B}$ are obtained by the critical-state principles $|J_{\perp}|\leq J_{c\perp}$, $|J_{\|}|\leq J_{c\|}$, appropriate $\vec{E}(\vec{J})$ laws and the Maxwell equations. Finite element based models are also available\cite{bosspri}, which allow to compute the critical state profiles for non-ideal geometries.

Here, we will show that Bean's simplest concept of opposing external field variations with the maximum current density is still valid for multicomponent situations; the sign selection for one dimensional problems will become a particular case of finding the adequate direction of $\vec{J}$ by way of a minimum principle. Variational approaches based on the minimization of the free energy for calculating the magnetic properties of type-II superconductors have been applied before. In Ref.\cite{navau} a numerical method is presented which allows computing the current distribution for the Meissner state in finite cylinders. Also, in a previous work\cite{badia} we showed that a powerful generalization of variational calculus, the optimal control (OC) theory\cite{oc}, provides a very convenient mathematical framework for critical-state problems in superconductors. Generally speaking, the OC tools may be fully exploited in physical theories that include limitations in the form of inequalities. In Ref.\cite{badia}, the magnetostatic energy was minimized under the restriction $|\vec{J}|\leq J_c$. That principle may only be applied to the initial magnetization curve, as hysteretic losses cannot be accounted by a thermodynamic equilibrium model. Nevertheless, it was suggested that a functional, most probably related to changes in the magnetic field vector should allow dealing with the full problem. In this letter we show that the OC tools may be used to predict the irreversible quasistationary evolution that takes place. Eventually, the theory will be applied to the experiments of a superconductor in rotating magnetic fields\cite{leblanc} and the magnetization collapse in crossed field measurements\cite{park}.

The basic relations of coarse-grained electrodynamics in the case of type-II superconductors read as follows. As time-dependent phenomena are involved,
one must incorporate Faraday's law $\mu_{0}\partial_{t}\vec{H}=-\nabla\times \vec{E}$ (we have used $\vec{B}=\mu_0\vec{H}$, which means that reversible magnetization is neglected in this work), as well as an appropriate $\vec{E}(\vec{J})$ characteristic for the superconductor, where
$\vec{J}$ is to be obtained from Amper\`e's law $\nabla\times \vec{H}=\vec{J}$.

We will assume a discretization scheme in which $\vec{H}_{\rm n}$
stands for the magnetic field intensity at the time layer $n\delta t$. This procedure permits posing the minimum principle as a tractable boundary value problem for ordinary differential equations. In order to gain physical insight, we will infer the CSM equations after considering some aspects of the more familiar eddy-current problem in normal metals ($\vec{E}=\rho \vec{J}$). The successive field profiles in a magnetic diffusion process may be obtained by the finite-difference expression $\mu_{0}(\vec{H}_{\rm n+1}-\vec{H}_{\rm n})/\delta t=-\rho \nabla\times(\nabla\times \vec{H}_{\rm n+1})$, which defines a differential equation for $\vec{H}_{\rm n+1}$. Notice that for each step one can identify it as the {\em stationarity} condition for the functional
\[
{\cal C}_{\rm M}[\vec{H}_{\rm n+1}]
=\mu_{0}\int_{\Omega}\! |\vec{H}_{\rm n+1}-\vec{H}_{\rm n}|^{2}
+\delta t\int_{\Omega}\! \vec{E}\cdot\vec{J}
\equiv \int_{\Omega}\! {\cal F}_{\rm n+1}\; ,
\]
where $\Omega$ stands for the sample's volume and the dependence of the second term on $\vec{H}_{\rm n+1}$ is implicitly assumed. In fact, the Euler-Lagrange equations which describe the stationarity of ${\cal C}_{\rm M}$, i.e.: 
$\partial{\cal F}_{\rm n+1}/\partial H_{\rm n+1,i}=\partial_{x_{\rm j}} [\partial{\cal F}_{\rm n+1}/\partial (\partial_{x_{\rm j}} H_{\rm n+1,i})]$ can be checked to produce the aforementioned expression. We call the readers' attention that ${\cal C}_{\rm M}$ should not be mistaken for the action in the classical theory of fields $S=\int\int\! {\cal L}\, dV\, dt$. As an additional advantage of using ${\cal C}_{\rm M}$, we get a clear physical picture of the underlying series of quasistationary processes. Notice that ${\cal C}_{\rm M}$ holds a compensation between a {\em screening} term and an {\em entropy production} term. In fact, under isothermal conditions one has $\dot{\cal S}=\vec{E}\cdot\vec{J}/T$. Thus, a perfect conductor would correspond to the limit $\dot{\cal S}\to 0\Rightarrow \vec{H}_{\rm n+1}\to\vec{H}_{\rm n}$. On the opposite side, non-conducting media would not allow the existence of screening currents (otherwise $\dot{\cal S}\to\infty$) and, thus, $\vec{H}_{\rm n+1}$ will be solely determined by the external source. In the case of type-II superconductors, the critical state arises from the flux flow characteristic, which, in the isotropic case, can be written as $E=\rho_{\rm f}(J-J_{c})$ (or 0 if $J<J_{c}$). Thus, the external drive variations are followed by diffusion towards equilibrium critical profiles in which $J$ equals $J_{c}$. If the relaxation time $\tau_{\rm f}$ may be neglected (or equivalently $\rho_{\rm f}\to\infty$) the superconductor will behave as a perfect conductor for $J\leq J_{c}$ and as a non-conducting medium for $J>J_{c}$. In the light of the previous discussion, the evolutionary critical state profiles can be either obtained by using Maxwell equations and a vertical $E(J)$ law or the principle: 

{\em In a type-II superconducting sample $\Omega$ with an initial
field profile $\vec{H}_{\rm n}$ and under a small change of the external drive, the new profile $\vec{H}_{\rm n+1}$ minimizes the functional}
\[
{\cal C}[\vec{H}_{\rm n+1}(\vec{x})]
=\frac{1}{2}\int_{\Omega}\! | \vec{H}_{\rm n+1} -
\vec{H}_{\rm n} |^{2} \; ,
\]
{\em for the boundary conditions imposed by the external source, and the constraint $\nabla\times\vec{H}_{\rm n+1}\in\Delta (\vec{H}_{\rm n+1},\vec{x})$.}

For simplicity, we will use the isotropic hypothesis $|\nabla\times\vec{H}_{\rm n+1}|\leq J_c(| \vec{H}_{\rm n+1}|)$ hereafter, i.e.: $\Delta$ is a disk. This will provide a nice agreement of our simulations and the experimental facts. However, anisotropy can be easily incorporated, for instance by choosing $\Delta$ to be an ellipse or a rectangle (DCSM case) oriented over different axes. Although isotropy would not seem to be justified according to the underlying physical mechanisms of flux depinning ($J_{c\perp}$) and cutting ($J_{c\|}$), it may be supported by other reasons. As a matter of fact, an average description seems adequate for highly twisted soft FLLs for which flux cutting phenomena are much more effective than for rotating rigid parallel sublattices\cite{sudbo}. 

In order to see how the OC machinery arises, let us consider an infinite slab of thickness $2a$ in a field parallel to the faces ($YZ$ plane) and take the origin of coordinates at the midplane. By virtue of the symmetry, we can restrict to the interval $0\leq x \leq a$. Along this work, we will use a Kim's model type\cite{kim} dependence of the critical current density $J_c(H)=J_{c_{0}}/(1+H/H_{0})$, which incorporates the microstructure dependent parameters $J_{c_{0}}$ and $H_{0}$. For convenience we will express $x$ in units of $a$, $H$ in units of $H_{0}$, and $J$ in units of $H_{0}/a$. Then we can state Amp\`ere's law, together with the critical current restriction in the following manner
\[
\frac{d\vec{H}_{\rm n+1}}{dx}=
\vec{f}(\vec{H}_{\rm n+1},\vec{u},x)=
\frac{\beta\vec{u}}{1+|\vec{H}_{\rm n+1}|} \; .
\]
Above we have introduced the dimensionless constant $\beta=J_{c_{0}}a/H_{0}$ and the so-called {\em control variable} $\vec{u}$, which is a vector within the unit disk $D$. Notice that, by construction, one has $\vec{u}\perp\vec{J}$. Thus, we have the {\em state equations} for the {\em state variables} $\vec{H}_{\rm n+1}(x)$.

Next, we require the minimization of the {\em functional} 
${\cal C}[\vec{H}_{\rm n+1}(x)]$ constrained by the state equations. Just in the manner of Ref.\cite{badia}, Pontryagin's maximum principle can be used to solve the OC problem. On defining the associated Hamiltonian
\[
{\cal H}=\vec{p}\cdot\vec{f}-\frac{1}{2}
(\vec{H}_{\rm n+1} -\vec{H}_{\rm n})^{2}\; ,
\]
the optimal solution (i.e.: functions $\vec{H}_{\rm n+1}^{*}(x)$ and $\vec{u}^{*}(x)$ minimizing ${\cal C}$ and satisfying the state equations) fulfils the Hamiltonian equations
\[
\frac{dH_{\rm n+1,i}^{*}}{dx}=f_{i} \quad , \quad 
\frac{dp_{i}^{*}}{dx}=H_{\rm n+1,i}^{*}-H_{\rm n,i}-p_{j}^{*}\frac{\partial f_j}{\partial H_{\rm n+1,i}^{*}}\; ,
\]
together with the maximum principle condition
\[
{\cal H}(\vec{H}_{\rm n+1}^{*},\vec{p}^{\,*},\vec{u}^{*})=
\max_{\vec{u}\in D}{\cal H}(\vec{H}_{\rm n+1}^{*},\vec{p}^{\,*},\vec{u})\; .
\]
In the case under consideration, the control variables must take the form $\vec{u}^{*}=\vec{p}^{\,*}/p^{*}$, and this leads to the system
\begin{mathletters}
\label{eqncanslab:all}
\begin{eqnarray}
\frac{dH_{\rm n+1,i}^{*}}{dx}&=&\frac{p_{i}^{*}}{p^{*}}\,\frac{\beta}{1+H_{\rm n+1}^{*}}
\label{eqncanslab:a}
\\[1ex]
\frac{dp_i^{*}}{dx}&=&H_{\rm n+1,i}^{*}-H_{\rm n,i}+
\frac{\beta p^{*}H_{{\rm n+1},i}^{*}}
{H_{\rm n+1}^{*}(1+H_{\rm n+1}^{*})^2} \; .
\label{eqncanslab:b}
\end{eqnarray}
\end{mathletters}

A part of the boundary conditions required to solve this system of differential equations is given by the external field values at the surface $\vec{H}_{\rm n+1}^{*}(1)=\vec{H}[1,(n+1)\delta t]$. The remaining boundary conditions will be supplied, at every instant, according to the particular situation: (i) If the new profile matches the old one at a point $0< x^{*}< 1$, i.e.: $\vec{H}_{\rm n+1}^{*}(x^{*})=\vec{H}_{\rm n}(x^{*})$, these are the extra boundary conditions. $x^{*}$ can be determined by additional equations derived from the minimum cost requirement. In fact, one can prove that a free final parameter $x^{*}$ leads to the algebraic condition ${\cal H}(x^{*})=0$. (ii) If the new profile holds a variation which reaches the center of the slab, the full arbitrariness of $\vec{H}_{\rm n+1}^{*}(0)$ supplies the so-called {\em transversality conditions} for the momenta: $\vec{p}^{\,*}(0)=0$.

Notice that the physical counterpart of the result $|\vec{u}^{*}(x)| =1$ is $|\vec{J}| =J_c(H)$. We should emphasize that this condition and the distribution rule for the components of $\vec{J}$ are determined by the selection of the control space $\Delta$. For instance, $|\vec{J}|$ is no more fixed when $\Delta$ is a rectangle. Instead, the optimality produces a vector leaning on the boundary, matching the evolution predicted by the DCSM.

Eventually, the critical state profiles will be solved by integration of the set of Eqs.(\ref{eqncanslab:all}). Below, we apply the method to the rotating and crossed field experiments. For definiteness, we choose $\beta=1$.

First, we consider the solutions for $\vec{H}_{\rm n+1}^{*}$ in a field cooled sample, which is subsequently subjected to a surface field rotation in the manner $\vec{H}_{S}(t)= H_{S}(0,\sin \alpha_{S},\cos\alpha_{S})$, where $\alpha_{S}\equiv\omega t$. On neglecting the equilibrium magnetization contribution, the slab holds a {\em nonmagnetic} initial state of constant profile $(0,0,H_{S})$. Successive profiles of the penetration field $\vec{H}_{\rm n+1}$ were obtained by means of Eqs.(\ref{eqncanslab:all}). Figure \ref{fig1} displays the main features of the calculated magnetization process in our system. During the initial stages of rotation the magnitude of $H$ (upper panel) is decreased towards the center of the slab in a {\em flux consumption} regime. Simultaneously, the angle of rotation $\alpha$ respect to $\vec{H}_{S}(0)$ follows a quasilinear penetration profile (lower panel). As rotation is continued the field modulus penetration curve develops a V-shape, which neatly defines a decoupling point $x_{0}$. Thereafter, the curve essentially freezes and the flux density modulus becomes stationary. On the other hand, the rotation angle variation is blocked in the range $0\leq x\leq x_{0}$. Further changes of $\vec{H}_{S}$ will only affect $\alpha (x)$ for  $x_{0}\leq x\leq 1$. In particular, after {\em decoupling} occurs, the external drive variations induce a conventional critical state behavior for the profile $\alpha (x)$. For instance, one can observe the expected effect of rotation reversal after one turn is completed (see the inset in the lower panel). Eventually, the outer region will be responsible for the hysteretic losses as the inner part contains an {\em inert} magnetic flux density distribution. 

The phenomenological matters described above have been experimentally observed by Cave and LeBlanc\cite{leblanc} and reported by Clem and P\'erez-Gonz\'alez\cite{clemgonzalez} from the theoretical point of view. However, we want to remark that DCSM model contains critical slopes for the field modulus and rotation angle, and the appearance of a decoupling point is somehow forced. Our variational principle allows showing that, even for the isotropic hypothesis, in which no such {\em a priori} condition is introduced, the optimal process itself produces the actual current distribution and generates decoupling. Thus, the behavior observed in Fig.\ref{fig1} is more related to the imposed boundary conditions than to the particular region $\Delta$ in use.

Next, we concentrate on the so called {\em magnetization collapse}, which can be observed in crossed field measurements. To our knowledge, Ref.\cite{park} displays a remarkable manifestation of this effect in high $T_c$ superconductors. We have calculated the field penetration profiles for a zero field cooled sample to which a constant excitation $H_{zS}$ is then applied. This is followed by cycling stages of the other field component on the surface $H_{yS}$. Fig.\ref{fig2} displays the predicted magnetization curves. As usual, we have defined $\vec{M}\equiv\langle \vec{H}(x)\rangle -\vec{H}_{S}$. The major loop (sequence OABC) displays the observed experimental features: (i)$M_y$ shows a nearly conventional CSM profile, except for the fact that the loop is not closed (see A and C). (ii) $M_z$ irreversibly {\em collapses} as $H_{yS}$ is cycled. Both effects may be easily explained in terms of the predicted penetration profiles. In the insets we show the evolution corresponding to the branch A $\to$ B. For illustration, we have also included a few profiles associated to the first field reversal steps in the branch B $\to$ C. Notice that $H_y$ follows the typical CSM pattern, whereas $\langle H_{z}\rangle$ continuously increases. Physically, this behavior must be related to the most effective mechanism for minimizing $\int |\vec{H}_{\rm n+1}-\vec{H}_{\rm n}|^{2}dx$. The current density component $J_z$ dedicated to reduce the imposed field variation $|H_{\rm n+1,y}-H_{\rm n,y}|^{2}$ is privileged. Then we have $J_y\simeq 0$ near the surface owing to the restriction on $|\vec{J}|$, and this leads to the flattening of $H_z$. We have also simulated a minor loop, which corresponds to a partial penetration regime. It is noteworthy that quite different behaviors can be observed depending on the applied field amplitudes.

In summary, we have presented a phenomenological critical state model that generalizes the minimal tool proposed by Bean to systems of twisted vortex configurations. Our theory may be used to understand the experimental features of rotating field experiments. Within the isotropic hypothesis, the model also provides a straightforward explanation of the observed magnetization collapse, for which a merely approximate justification was available\cite{voloshin}. Although this work has been developed for an isotropic relation of the kind $J \leq J_{c0}/(1+H/H_{0})$, several extensions may be implemented if dictated by the physics of the problem. These include the effect of equilibrium magnetization by means of an appropriate $B(H)$ relation, the selection of the model $J_{c}(H)$ and the use of anisotropic control spaces. Another important issue would be the incorporation of time relaxation effects by means of a finite flux flow resistivity. This can be accomplished by using the entropy production term $\vec{E}\cdot(\vec{J}-\vec{J}_{c})$ in the functional.

The authors acknowledge financial support from Spanish CICYT (project MAT99-1028) and from DGICYT (project DGES-PB96-0717).

%
%
%%%%%%%%%%%%%%%%%%%%%%%%%%%%%%%%%%%%%%%%%%%%%%%%%%%%%%%%%%%%%%%%
%%%%%%%%%%%%%             REFERENCES             %%%%%%%%%%%%%%%
%%%%%%%%%%%%%%%%%%%%%%%%%%%%%%%%%%%%%%%%%%%%%%%%%%%%%%%%%%%%%%%%
%
%

%
%
%
%%%%%%%%%%%%%%%%%%%%%%%%%%%%%%%%%%%%%%%%%%%%%%%%%%%%%%%%%%%%%%%%
%%%%%%%%%%%%%%%%%       FIGURES      %%%%%%%%%%%%%%%%%%%%%%%%%%%
%%%%%%%%%%%%%%%%%%%%%%%%%%%%%%%%%%%%%%%%%%%%%%%%%%%%%%%%%%%%%%%%
%
%
\begin{figure}
\caption[rotating field]{
Magnetic field penetration profiles for a field cooled slab under successive rotation steps for the surface vector $\vec{H}_{S}$. Panel (a) displays the modulus consumption towards the center of the slab ($x=0$) as well as the tendency towards a stationary frozen V-shape. Panel (b) displays the rotation angle $\alpha$ respect to the initial constant profile. Subsequent to the appearance of the decoupling point $x_0$, which has been marked on both graphs, the evolution restricts to the range $x_{0}<x<1$. The inset shows the calculated angle profiles upon rotation reversal. $H$ has been used in units of $H_0$, $x$ in units of $a$ and $\alpha$ is given in radians.
\label{fig1}}
\end{figure}
\vspace{1cm}
\begin{figure}
\caption[crossed field]{
Evolution of the magnetization components in a simulated crossed field experiment for a zero field cooled superconducting slab. Subsequent to the application of a constant surface field $H_{zS}$, the other component $H_{yS}$ was cycled either in a major loop (sequence OABC) or minor loop. Several magnetic field profiles, corresponding to the magnetization process have been included in the insets. Full symbols have been used for the curves corresponding to the points A and B, continuous lines for a selection of intermediate profiles and dashed lines for the initial steps in the branch B$\to$C. $H_y$ is plotted within the axis range (-0.6,0.6), $H_z$ within (0.25,0.65), and $x$ for (0,1). All the quantities are in dimensionless units as defined in the text.
\label{fig2}}
\end{figure}
\end{document}